\documentclass[11pt,a4paper]{article}
\usepackage{eurosym}
\usepackage[utf8]{inputenc}
\usepackage{amsmath}
\usepackage{amssymb}
\usepackage{amsfonts}
\usepackage{geometry}
\usepackage{cite}
\usepackage{graphicx}

\newtheorem{theorem}{Theorem}

\begin{document}

\title{\textbf{Geometric classification of Noether invariants of the
Time-Dependent damped Duffing Oscillator}}

\author{%
Michael Tsamparlis$^{1,2,3,a,b)}$ and Aniekan Ukpong$^{2,3,c)}$ \\[6pt]
$^{1}$\textit{Faculty of Physics, Department of
Astronomy-Astrophysics-Mechanics,}\\
\textit{University of Athens, Panepistemiopolis, Athens 157 83, Greece} \\[4pt]
$^{2}$\textit{NITheCS, National Institute for Theoretical and Computational
Sciences,}\\
\textit{Pietermaritzburg 3201, KwaZulu-Natal, South Africa} \\[4pt]
$^{3}$\textit{TCCMMP, Theoretical and Computational Condensed Matter and
Materials Physics Group,}\\
\textit{School of Agriculture and Sciences,}\\
\textit{University of KwaZulu-Natal, Pietermaritzburg 3201, South Africa} \\[4pt]
$^{a)}$Author to whom correspondence should be addressed. \\
$^{b)}$Email: mtsampa@phys.uoa.gr \\
$^{c)}$Email: ukponga@ukzn.ac.za
}
\date{}
\maketitle

\begin{abstract}
We present a systematic geometric classification of the Noether point
symmetries of time-dependent conformal Lagrangians of the form $L =
A(t)\left(\frac{1}{2}g_{ij}\dot{x}^{i}\dot{x}^{j} - V(x)\right)$, whose
Euler--Lagrange equations describe dynamical systems with time-dependent
linear damping. The classification is carried out by coupling the temporal
parameters of the conformal factor $A(t)$ with the homothetic algebra of the
kinetic metric $g_{ij}$ through a set of compatibility and integrability
conditions. The central result is a main theorem which determines the
Noether generators, the associated gauge functions and the resulting
invariants. The theorem splits naturally into two cases according to whether
the quantity $D - 2\psi$ is non-zero or zero, with further subcases
depending on whether the homothetic vectors are gradient or non-gradient.
The theorem imposes no restrictions on $A(t)$, $g_{ij}$ or $V(x)$, and is
therefore fully general and covariant. We apply the theorem to the
time-dependent damped Duffing oscillator (DDO) in Euclidean space $E^{3}$
with $A(t) = e^{\gamma t}$ and potential $V = \frac{1}{2}a\,r^{2} + \frac{%
\lambda}{4}r^{4}$. For the damped harmonic oscillator ($\lambda = 0$) the
theorem reproduces the classical Bateman invariant as a special case. For
the anharmonic case we obtain a new exact time-dependent quadratic
invariant, valid when the linear stiffness satisfies $a = 2\gamma^{2}/9$.
Each invariant is verified explicitly by direct differentiation along the
equations of motion. The new invariant provides a rare example of a global
conservation law for a fully nonlinear dissipative system, and can be used
for the integration and quantization of the DDO.
\end{abstract}

Keywords: Time-dependent conformal Lagrangians, Noether symmetries,
Homothetic motions, Duffing Oscillator, Anharmonic oscillator, time
dependent invariants.

\section{Introduction}

The identification of exact invariants and conservation laws remains a
fundamental pillar in the analysis of non-linear dynamical systems. While
conservative frameworks enjoy well-established geometric properties,
dissipative systems present severe structural challenges due to the explicit
symmetry-breaking introduced by friction terms. Historically, the classical
dissipative system, the damped harmonic oscillator (DHO), has been modeled
through auxiliary variable formulations which determines the time Ermakov -
Lewis invariant. This has been used in various ways e.g. for the solution of
the dynamical equations or as a Hamiltonian in the quantization of the DHO ~%
\cite{Bateman1931, Kanai1950,Lewis1968,MullerKirsten1979, Margheri2014,
Margheri2020, Kristiansen2024}. However, in more complex systems the
determination of invariants with this method is not available and one has to
work either by approximation methods, simplifications or an ansatz. One such
example is the anharmonic oscillator, or Duffing oscillator (DO) a holonomic
system with potential $V(r)=ar^2 +\lambda r^4$. The dynamical equations for
the undamped and the unforced case of the DO have been integrated in terms
of the Jacobi elliptic function cn \cite{Book,Salas2021}. For the damped and
the forced case this is not possible and one looks for approximate solutions
using various methods \cite{Salas2020A}.

In the present work we are not looking at the dynamical equations but we are
looking for the determination of invariants of dynamical systems with time
dependent linear damping. Obviously Noether point symmetries offer an
elegant covariant methodology to address this point.

As a consequence of the above we consider a general conformal Lagrangian of
the form:%
\begin{equation}
L = A(t)\left( \frac{1}{2}g_{ij}\dot{x}^{i}\dot{x}^{j} - V(x) \right).
\label{Lagrangian}
\end{equation}
where $g_{ij}$ is the kinetic metric defined by the kinetic energy (not the
metric of the space where motion takes place) and $V(x)$ is a general
potential. We then discuss the geometric classification of Noether point
symmetries of this type of Lagrangian. The reason for choosing this form of
Lagrangian is that the Euler Lagrange equations are: 
\begin{equation}
\ddot{x}^{i}+(\ln A)_{,t}\dot{x}^{i}+g^{ij}V_{,j} =0.  \label{EL_equations}
\end{equation}
These are the dynamical equations for a system with time dependent linear
damping term. In particular these equations generate the time dependent
damped DO (DDO) if one considers the Euclidean metric and the potential $%
V(r)=ar^2 +\lambda r^4$. Earlier applications of this approach on the damped
harmonic oscillator (DHO) and other damped systems can be found in ~\cite%
{Tsamparlis2026A}.

Lagrangian (\ref{Lagrangian}) contains elements of two different natures 
\newline
a. Geometric elements, which are the conformal function $A(t)$ and the
metric $g_{ij}$ \newline
b. Dynamical element, which is the potential specifying the dynamical
equations.\newline
As a result the Noether symmetries of this type of Lagrangian relate the
invariants of the dynamical system with the geometric elements $A(t),g_{ij}$
~\cite{MeAndroGRG2010, Tsamparlis2026A}. This result applies to other type
of Lagrangians, for example for autonomous Lagrangians and Lagrangians with
a time dependent separable potential of the form $\omega(t)V(x)$ \cite%
{LeoAndroME2017}.

The purpose of the present work is to establish a rigorous main theorem
(Theorem (\ref{Main_theorem})) which classifies the Noether invariants of
Lagrangians of the form (\ref{Lagrangian}). The classification is realized
by means of a coupling between temporal parameters and the homothetic
algebra of the kinetic metric through a series of compatibility and
integrability conditions. It is found that one has to consider two main
cases. Case I in which the quantity $D-2 \psi\ne{0}$ and case II in which $%
D-2 \psi=0$. Each case has subcases depending on the vectors of the
homothetic algebra being gradient or non-gradient.

We demonstrate the application of the Theorem to the case of the DDO defined
by the potential:%
\begin{equation}
V(x)=\frac{1}{2}ax^{i}x_{i}+\frac{\lambda }{4}(x^{i}x_{i})^{2}.
\label{Potential}
\end{equation}%
The parameter $a$ controls the linear stiffness and the parameter $\lambda $
controls the non-linearity of the DDO. Physically the equation of motion of
the DDO can be seen as describing the oscillations of a mass attached to a
nonlinear spring and a linear damper. The restoring force provided by the nonlinear spring is 
$ax^{j}+\lambda (x^{i}x_{i})x^{j}$.  Depending on the
values $\lambda >0,<0$ one considers the hardening and the softening spring.
Detailed information on the damped DO can be found in \cite{Book}.

The reason for choosing this system is that it is non-linear and, in the
forced case, exhibits chaotic behavior; if we obtain a genuine explicit
Noether invariant for the unforced system, it can be used to ease the study
of the behaviour of the system. Furthermore, because
when $\lambda=0$ the potential V(x) reduces to the potential of the damped
harmonic oscillator (DHO), this study will produce as a byproduct the
invariants of that classical non-linear system. Indeed as it will be shown
one obtains new invariants for the DHO which include the classic Bateman
invariant as a special case.

The structure of the paper is as follows. In section \ref{sec:Main_therem1}
we state the main Theorem. In section \ref{sec:Aexp-central} we consider the
DDO and determine the Noether quantities. The results concern two cases a.
The DHO ($\lambda=0$) which gives the Noether quantities of the DHO and b. A
new non-trivial invariant of the DDO. In section \ref{sec:verification} we
verify the invariants both analytically and numerically. Finally in
section \ref{Conclusion} we draw our conclusions. The proof of the theorem
is in the appendix.

\section{The Noether symmetries of time dependent conformal Lagrangian}

\label{sec:Main_therem1}

\begin{theorem}
\label{Main_theorem} Consider a dynamical system described by the
time-dependent Lagrangian 
\begin{equation}
L(t,x^{k},\dot{x}^{k})=A(t)\left[ \frac{1}{2}g_{ij}(x^{k})\dot{x}^{i}\dot{x}%
^{j}-V(x^{k})\right] ,  \label{eq:GEN.1}
\end{equation}
with $A(t)\neq 0$  and in Case I we require $(\ln A)_{,t}\neq 0$. The Noether point symmetry generators are of the form 
\begin{equation}
X=\xi ^{0}(t)\partial _{t}+\xi ^{i}(t,x^{k})\partial _{x^{i}},
\end{equation}
where:

\begin{enumerate}
\item The spatial component is 
\begin{equation}
\xi ^{i}(t,x)=T(t)Y^{i}(x),  \label{THM.2}
\end{equation}
where $Y^{i}(x)$ is an element of the homothetic algebra of the metric,
satisfying $\mathcal{L}_{Y}g_{ij}=2\psi g_{ij}$ ($\psi =0$ for a Killing
vector, $\psi =1$ for a homothetic vector), and $T(t)$ is the temporal
scaling parameter.

\item The potential satisfies the spatial identity 
\begin{equation}
(Y(V))_{,i}+DV_{,i}+mY_{i}=0,  \label{eq:constr_V}
\end{equation}
where the constants $D,m$ are defined by 
\begin{equation}
\frac{(A(t)\xi ^{0}(t))_{,t}}{A(t)T(t)}=D,  \label{eq:def1_D}
\end{equation}
\begin{equation}
\frac{(A(t)T_{,t}(t))_{,t}}{A(t)T(t)}=m.  \label{eq:def1_m}
\end{equation}

\item The temporal scaling parameter $T(t)$ satisfies 
\begin{equation}
T_{,tt}+(\ln A)_{,t}T_{,t}-mT(t)=0.  \label{THM.3}
\end{equation}
\end{enumerate}

\noindent We distinguish two major cases.

\medskip \noindent\textbf{Case I: $D-2\psi \neq 0$.} In this case $(\ln
A)_{,t}\neq 0$, and the temporal component of the generator is 
\begin{equation}
\xi ^{0}(t)=\frac{(D-2\psi )T(t)}{2(\ln A)_{,t}}.  \label{eq:master_xi0_sol1}
\end{equation}
The conformal factor and the temporal scaling satisfy the constraint 
\begin{equation}
\frac{d}{dt}\left[ \frac{(D-2\psi )T(t)}{(\ln A)_{,t}}\right] -(D-2\psi
)T(t)=4\psi T(t).  \label{eq:final_constraint1}
\end{equation}
Two subcases arise.

\medskip \noindent\textit{Case I-A: $T_{,t}\neq 0$.} The vector $Y^{i}$ is a
gradient vector with generating function $M(x)$, $Y^{i}=M^{,i}$. The
constants $D,m$ are determined from 
\begin{equation}
Y(V)+DV+mM=C_{1},  \label{NL.10c1}
\end{equation}
where $C_{1}$ is a constant. The Noether gauge function is 
\begin{equation}
f(t,x)=A(t)T_{,t}M(x)-C_{1}\int A(t)T(t)\,dt.  \label{NC.10c2}
\end{equation}

\medskip \noindent\textit{Case I-B: $T_{,t}=0$, $T=a_{1}$.} The vector $%
Y^{i} $ may be gradient or non-gradient. The constant $D$ is determined from 
\begin{equation}
Y(V)+DV=C_{2},  \label{NL.10d}
\end{equation}
the temporal component is 
\begin{equation}
\xi ^{0}(t)=\frac{a_{1}(D-2\psi )}{2(\ln A)_{,t}},
\label{eq:master_xi0_sol21}
\end{equation}
and the gauge function is 
\begin{equation}
f(t)=-C_{2}a_{1}\int A(t)\,dt.  \label{NL.10c3}
\end{equation}

\medskip \noindent\textbf{Case II: $D-2\psi =0$.} In this case $\xi ^{0}=0$,
and the quadratic constraint forces $\psi =0$. Hence $Y^{i}$ is a Killing
vector (not necessarily gradient) and $D=0$. The potential satisfies 
\begin{equation}
(Y(V))_{,i}+mY_{i}=0.  \label{NL.10c1_D2}
\end{equation}
Two subcases arise.

\medskip \noindent\textit{Case II-A: $Y^{i}$ is a gradient KV, $Y_{i}=M_{,i}$%
.} The constant $m$ is determined from 
\begin{equation}
Y(V)+mM=C_{1},  \label{NL.10c1_D2_grad}
\end{equation}
and $T(t)$ satisfies \eqref{THM.3}. The gauge function is given by %
\eqref{NC.10c2}.

\medskip \noindent\textit{Case II-B: $Y_{i}$ is not a gradient KV.} Then $%
T=a_{1}=\text{constant}$, $m=0$, and the potential satisfies 
\begin{equation}
Y(V)=C_{2}.  \label{eq:constr_V_D2_nongrad}
\end{equation}
The gauge function is 
\begin{equation}
f(t)=-C_{2}a_{1}\int A(t)\,dt.  \label{eq:fin_T0_D2}
\end{equation}

\medskip \noindent In all cases the Noether integral is 
\begin{equation}
I=A(t)\left( TY_{i}\dot{x}^{i}-\xi ^{0}(t)\left( \frac{1}{2}\dot{x}^{i}\dot{x%
}_{i}+V(x)\right) \right) -f(t,x).  \label{THM.7.1}
\end{equation}
In Case II, since $\xi ^{0}=0$, the invariant reduces to $I=A(t)T Y_{i}\dot{x%
}^{i}-f(t,x)$, which is linear in the velocities.
\end{theorem}

The proof of the theorem is given in the appendix.

\section{The Damped Duffing Oscillator (DDO)}

\label{sec:Aexp-central}

We apply Theorem \ref{Main_theorem} to the DDO with the scaling $A(t) =
e^{\gamma t}$ ($\gamma \neq 0$): 
\begin{equation}
L = e^{\gamma t}\left[ \frac{1}{2}\delta_{ij}\dot{x}^{i}\dot{x}^{j} - \left( 
\frac{a}{2}x_{i}x^{i} + \frac{\lambda }{4}(x_{i}x^{i})^{2} \right) \right].
\label{eq:DDO_Lagrangian}
\end{equation}
The associated Euler--Lagrange equations yield the Duffing oscillator: 
\begin{equation}
\ddot{x}^{i} + \gamma \dot{x}^{i} + (a + \lambda r^{2})x^{i} = 0,
\label{eq:DDO_EOM}
\end{equation}
where $r^{2} = x_{i}x^{i}$. For $a=-1$, $\lambda=1$ this simplifies to the unforced
Holmes--Duffing model, and for $\lambda=0$ it reduces to the standard damped
harmonic oscillator (DHO).

\subsection{Homothetic Algebra and Potential Conditions}

The homothetic algebra of $E^{3}$ comprises:

\begin{enumerate}
\item \textbf{Gradient Killing vectors (translations):} $Y^{i} =
\partial_{i} $ with generating function $M(x) = x_i$.

\item \textbf{Non-gradient Killing vectors (rotations):} $Y^{i} =
\varepsilon^{ijk}x_j\partial_k$.

\item \textbf{Gradient Homothetic vector (dilatation):} $Y^{4} =
x^{i}\partial_{i}$ with $\psi = 1$ and generating function $M(r) = \frac{1}{2%
}r^{2}$.
\end{enumerate}

Evaluating these vector fields on $V(r) = \frac{1}{2}ar^{2} + \frac{\lambda}{%
4}r^{4}$ gives:

\begin{itemize}
\item \textbf{Gradient KVs (translations):} $Y(V) = (a + \lambda r^{2})x^{i}$%
.

\item \textbf{Non-gradient KVs (rotations):} $Y(V) = 0$.

\item \textbf{Gradient HV (dilatation):} $Y^{4}(V) = ar^{2} + \lambda r^{4}
= 2V + \frac{1}{2}\lambda r^{4}$.
\end{itemize}

\subsection{Case I-A: $T_{,t}\neq 0$}

In Case I, with $(\ln A)_{,t} = \gamma$, the temporal scaling parameter
satisfies 
\begin{align}
T_{,tt} + \gamma T_{,t} - m T(t) &= 0,  \label{eq:T_trans_exp} \\
T_{,t} &= \gamma \left(\frac{D+2\psi }{D-2\psi }\right)T.
\label{eq:T_const_exp}
\end{align}
Solving this coupled system yields 
\begin{align}
T(t) &= T_{0}e^{\frac{D+2\psi }{D-2\psi }\gamma t},  \label{T_exp} \\
m &= \frac{2D(D+2\psi )}{(D-2\psi )^{2}}\gamma ^{2}.  \label{m_exp}
\end{align}
The temporal component of the generator is 
\begin{equation}
\xi ^{0}(t) = \frac{(D-2\psi )T(t)}{2\gamma},  \label{eq:xi0_DDO}
\end{equation}
and the gauge function is 
\begin{equation}
f(t,x) = e^{\gamma t}T_{,t}M(x) - C_{1}\int e^{\gamma t}T(t)\,dt.
\label{eq:f_DDO}
\end{equation}

For the dilatation vector ($Y^{4}$, $\psi=1$, $M=\frac{1}{2}r^{2}$),
matching coefficients in the potential identity 
\begin{equation}
Y(V) + DV + mM = C_{1}  \label{eq:pot_identity_DDO}
\end{equation}
yields three valid parameter layouts.

\subsubsection{Branch 1: $D \neq -2, -4$ (DHO, $\protect\lambda = 0$)}

For $D \neq -2, -4$, the potential identity forces $\lambda = 0$ and 
\begin{equation}
a = -\frac{2D}{(D-2)^{2}}\gamma^{2}, \qquad V(r) = -\frac{D\gamma^{2}}{%
(D-2)^{2}}r^{2}.  \label{eq:U_HV_exp}
\end{equation}
The constant $C_{1} = 0$. The generator components are 
\begin{equation}
\xi^{0}(t) = \frac{D-2}{2\gamma}T(t), \qquad f(t,x) = \frac{D+2}{2(D-2)}%
\gamma r^{2}e^{\gamma t}T(t).
\end{equation}
Substituting into the Noether integral \eqref{THM.7.1} gives 
\begin{equation}
I = T_{0}e^{\frac{2D\gamma}{D-2}t}\left( x_{i}\dot{x}^{i} - \frac{D-2}{%
4\gamma}\dot{x}^{i}\dot{x}_{i} - \frac{\gamma}{D-2}r^{2} \right).
\label{eq:I_HV_exp}
\end{equation}
This matches equation (32) of \cite{Tsamparlis2026A} for the damped harmonic
oscillator (up to an overall sign, which is a convention).

\subsubsection{Branch 2: $D = -2$ (Bateman invariant, DHO)}

For $D = -2$, the exponent in \eqref{T_exp} vanishes, so $T = T_{0}$ is
constant and $m = 0$. The potential is the DHO potential $V(r) = \frac{1}{2}%
ar^{2}$ with $a$ arbitrary, and $C_{1} = 0$. The temporal component is 
\begin{equation}
\xi^{0} = -\frac{2T_{0}}{\gamma},
\end{equation}
and the gauge function vanishes ($T_{,t} = 0$, $C_{1} = 0$). The Noether
integral \eqref{THM.7.1} gives 
\begin{equation}
I = e^{\gamma t}T_{0}\left( \dot{x}^{i}\dot{x}_{i} + \gamma x_{i}\dot{x}^{i}
+ ar^{2} \right).  \label{eq:I_D=-2_exp}
\end{equation}
This is the classical Bateman invariant for the damped harmonic oscillator 
\cite{Bateman1931,Kerner1958}.

\subsubsection{Branch 3: $D = -4$ (new invariant, DDO)}

For $D = -4$, the potential identity forces 
\begin{equation}
a = \frac{2}{9}\gamma^{2}, \qquad m = \frac{4}{9}\gamma^{2}, \qquad C_{1} =
0,
\end{equation}
and the potential is 
\begin{equation}
V(r) = \frac{2\gamma^{2}}{9}r^{2} + \frac{\lambda}{4}r^{4}.
\label{eq:V_branch3}
\end{equation}
The temporal scaling parameter and the temporal component are 
\begin{equation}
T(t) = T_{0}e^{\gamma t/3}, \qquad \xi^{0}(t) = -\frac{3}{\gamma}T(t),
\end{equation}
and the gauge function is 
\begin{equation}
f(t,x) = \frac{\gamma}{6}r^{2}e^{\gamma t}T(t).
\end{equation}
The generator is 
\begin{equation}
X = T\left( -\frac{3}{\gamma}\partial_{t} + x^{i}\partial_{i} \right).
\label{eq:X_branch3}
\end{equation}
Substituting into the Noether integral \eqref{THM.7.1} gives the new exact
invariant 
\begin{equation}
I = T_{0}e^{\frac{4}{3}\gamma t}\left( x_{i}\dot{x}^{i} + \frac{3}{2\gamma}%
\dot{x}^{i}\dot{x}_{i} + \frac{\gamma}{6}r^{2} + \frac{3\lambda}{4\gamma}%
r^{4} \right).  \label{eq:I_HV_DD=-4_ANO}
\end{equation}
This is the main result of the paper: a new exact time-dependent quadratic
invariant for the anharmonic DDO, valid when the linear stiffness satisfies $%
a = \frac{2}{9}\gamma^{2}$. The parameter $\gamma$ has dimensions of inverse
time, and the condition $a = \frac{2}{9}\gamma^{2}$ therefore relates a
stiffness (dimensions of frequency squared) to a damping rate squared.

\subsection{Case I-B: $T_{,t} = 0$ ($T = a_{1}$)}

In Case I-B, $T = a_{1}$ is constant, and the constraint %
\eqref{eq:final_constraint1} with $(\ln A)_{,t} = \gamma$ gives 
\begin{equation}
\frac{d}{dt}\left[\frac{1}{\gamma}\right] = 0 = \frac{D+2\psi}{D-2\psi},
\end{equation}
so $D = -2\psi$. For the dilatation vector ($\psi = 1$), this gives $D = -2$%
, which is Branch 2 (Bateman). For the rotations ($\psi = 0$), this gives $D
= 0$, which is actually Case II (since $D - 2\psi = 0$). We therefore treat
the rotations in Section \ref{sec:caseII_DDO} below.

\subsection{Case II: $D - 2\protect\psi = 0$}

\label{sec:caseII_DDO}

In Case II, $\xi^{0} = 0$, $\psi = 0$, and $D = 0$. The potential satisfies 
\begin{equation}
(Y(V))_{,i} + mY_{i} = 0.  \label{eq:ConV_DDO}
\end{equation}

\subsubsection{Case II-A: Gradient KV (translations)}

For the translations $Y^{i} = \partial_{i}$ ($\psi = 0$, $Y(V) = (a +
\lambda r^{2})x^{i}$), the potential identity integrates to 
\begin{equation}
Y(V) + mM = C_{1}.
\end{equation}
With $M = x_i$ (generating function of the translation), this gives 
\begin{equation}
(a + \lambda r^{2})x^{i} + mx^{i} = C_{1},
\end{equation}
which forces $\lambda = 0$, $C_{1} = 0$, and $m = -a$. The potential is the
DHO potential $V = \frac{1}{2}ar^{2}$, and $T(t)$ satisfies 
\begin{equation}
T_{,tt} + \gamma T_{,t} + aT(t) = 0.  \label{eq:T_caseII_trans}
\end{equation}
The gauge function and the invariant follow from \eqref{NC.10c2} and %
\eqref{THM.7.1}. Since these invariants are well known \cite%
{Choudhuri2008,Tsamparlis2026A}, we do not reproduce them here.

\subsubsection{Case II-B: Non-gradient KV (rotations)}

For the rotations $Y^{i} = \varepsilon^{ijk}x_{j}\partial_{k}$ ($\psi = 0$, $%
Y(V) = 0$), the potential identity gives $C_{2} = 0$. With $T = a_{1}$
constant, $m = 0$, and $f = 0$, the invariant \eqref{THM.7.1} reduces to 
\begin{equation}
I = a_{1}e^{\gamma t}\varepsilon^{ijk}x_{j}\dot{x}_{k}.
\label{eq:I_rotation}
\end{equation}
This is the angular momentum $L_{i} = \varepsilon^{ijk}x_{j}\dot{x}_{k}$
scaled by the factor $e^{\gamma t}$, holding for \textbf{all} central
potentials $V(r)$.

\section{Verification of the Invariants}

\label{sec:verification}

We verify explicitly the invariants (\ref{eq:I_HV_exp}), (\ref{eq:I_D=-2_exp}), and (\ref{eq:I_HV_DD=-4_ANO}).

\subsubsection{Verification of Branch 1}

Define 
\begin{equation}
\Phi _{1}=x_{i}\dot{x}^{i}-\frac{D-2}{4\gamma }\dot{x}^{i}\dot{x}_{i}-\frac{%
\gamma }{D-2}r^{2},
\end{equation}%
so that $I=T_{0}e^{\frac{2D\gamma }{D-2}t}\Phi _{1}$. We need to prove $\dot{%
\Phi}_{1}=-\frac{2D\gamma }{D-2}\Phi _{1}$. Using \eqref{eq:DDO_EOM} with $%
\lambda =0$ and $a=-\frac{2D\gamma ^{2}}{(D-2)^{2}}$ we compute:%
\begin{equation}
\dot{\Phi}_{1}=\frac{D}{2}\dot{x}^{i}\dot{x}_{i}+\left( -\gamma +\frac{(D-2)a%
}{2\gamma }-\frac{2\gamma }{D-2}\right) x_{i}\dot{x}^{i}-ar^{2}.
\end{equation}%
With $a=-\frac{2D\gamma ^{2}}{(D-2)^{2}}$, the coefficient of $x_{i}\dot{x}%
^{i}$ simplifies to $-\frac{2D\gamma }{D-2}$, and $-a=\frac{2D\gamma ^{2}}{%
(D-2)^{2}}$. Substituting we find:
\begin{equation}
\dot{\Phi}_{1}=\frac{D}{2}\dot{x}^{i}\dot{x}_{i}-\frac{2D\gamma }{D-2}x_{i}%
\dot{x}^{i}+\frac{2D\gamma ^{2}}{(D-2)^{2}}r^{2}=-\frac{2D\gamma }{D-2}\Phi
_{1}.
\end{equation}%
Therefore $dI/dt=0$. \hfill $\square $

\subsubsection{Verification of Branch 2 (Bateman invariant)}

Define
\begin{equation}
\Phi _{2}=\dot{x}^{i}\dot{x}_{i}+\gamma x_{i}\dot{x}^{i}+ar^{2},
\end{equation}
so that $I=e^{\gamma t}T_{0}\Phi _{2}$. We need to prove $\dot{\Phi}_{2}
=-\gamma \Phi _{2}$. Using \eqref{eq:DDO_EOM} with $\lambda =0$:
\begin{align}
\frac{d}{dt}(\dot{x}^{i}\dot{x}_{i}) &= -2\gamma \dot{x}^{i}\dot{x}_{i}-2ax_{i}\dot{x}^{i}, \\
\frac{d}{dt}(x_{i}\dot{x}^{i}) &= \dot{x}^{i}\dot{x}_{i}-\gamma x_{i}\dot{x}^{i}-ar^{2}, \\
\frac{d}{dt}(r^{2}) &= 2x_{i}\dot{x}^{i}.
\end{align}
Substituting,
\begin{equation}
\dot{\Phi}_{2}=-\gamma \dot{x}^{i}\dot{x}_{i}-\gamma ^{2}x_{i}\dot{x}%
^{i}-\gamma ar^{2}=-\gamma \Phi _{2}.
\end{equation}
Therefore $dI/dt=0$.\hfill $\square$

\subsubsection{Verification of Branch 3 (new invariant)}

Define 
\begin{equation}
\Phi _{3}=x_{i}\dot{x}^{i}+\frac{3}{2\gamma }\dot{x}^{i}\dot{x}_{i}+\frac{%
\gamma }{6}r^{2}+\frac{3\lambda }{4\gamma }r^{4},
\end{equation}%
so that $I=T_{0}e^{\frac{4}{3}\gamma t}\Phi _{3}$. We need to prove $\dot{%
\Phi}_{3}=-\frac{4}{3}\gamma \Phi _{3}$. Using \eqref{eq:DDO_EOM} with $a=%
\frac{2}{9}\gamma ^{2}$ we calculate:
\begin{equation}
\dot{\Phi}_{3}=-2\dot{x}^{i}\dot{x}_{i}+\left( -\frac{2\gamma }{3}-\frac{3a}{%
\gamma }\right) x_{i}\dot{x}^{i}-ar^{2}-\lambda r^{4}.
\end{equation}%
With $a=\frac{2}{9}\gamma ^{2}$, the coefficient of $x_{i}\dot{x}^{i}$
becomes $-\frac{4\gamma }{3}$, and $-a=-\frac{2}{9}\gamma ^{2}$. Replacing we find:
\begin{equation}
\dot{\Phi}_{3}=-2\dot{x}^{i}\dot{x}_{i}-\frac{4\gamma }{3}x_{i}\dot{x}^{i}-%
\frac{2}{9}\gamma ^{2}r^{2}-\lambda r^{4}=-\frac{4\gamma }{3}\Phi _{3}.
\end{equation}%
Therefore $dI/dt=0$.  \hfill $\square $

\subsection{Numerical verification of the new invariant}

\label{sec:numerical}

\begin{figure}[htbp]
\centering
\includegraphics[width=\textwidth]{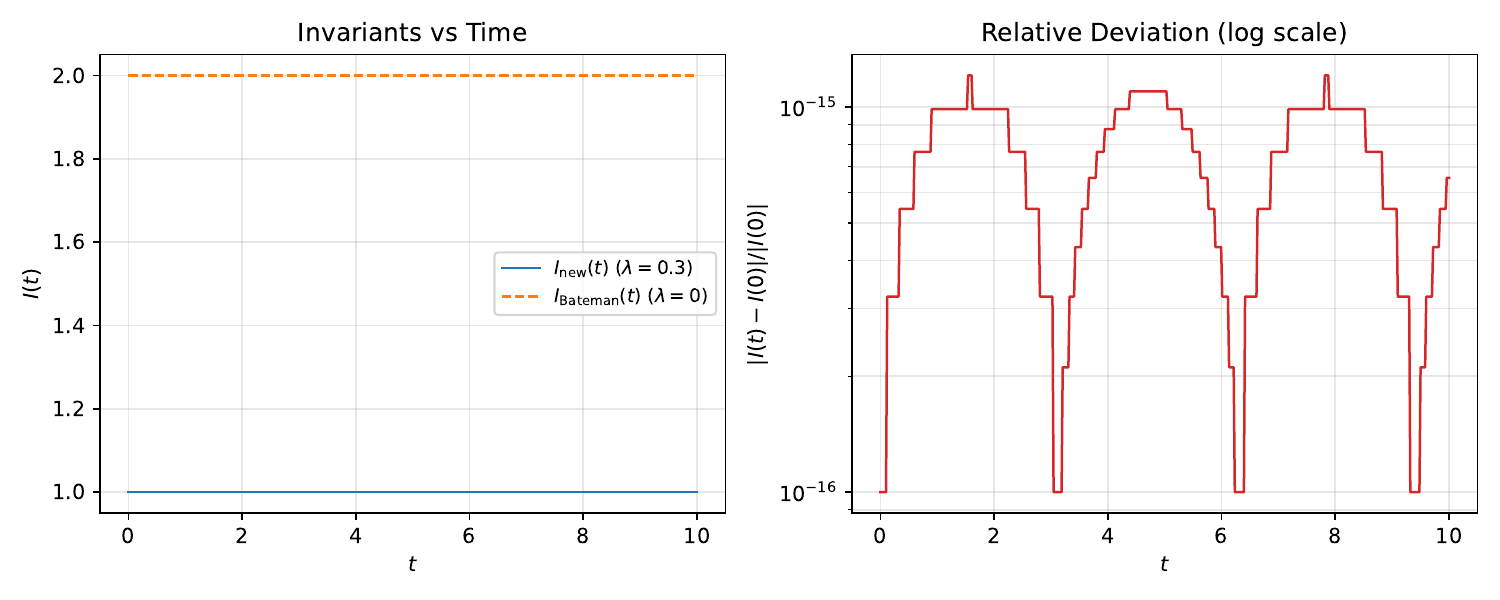}
\caption{Left: the new invariant $I_{\mathrm{new}}(t)$ for the anharmonic
DDO ($\lambda = 0.3$) and the Bateman invariant $I_{\mathrm{Bateman}%
}(t)$ for the DHO ($\lambda = 0$), both with $\gamma = 0.5$
and $a = 2\gamma^{2}/9$. Right: relative deviation $%
|I(t)-I(0)|/|I(0)|$ on a logarithmic scale. Both invariants are conserved to
machine precision, confirming the analytical results of Sections \protect\ref%
{sec:Aexp-central} and \protect\ref{sec:verification}.}
\label{fig:invariant}
\end{figure}

To complement the analytical verification of Section \ref{sec:verification},
we integrated the equations of motion \eqref{eq:DDO_EOM} numerically for the
anharmonic case with $\gamma = 0.5$, $a = 2\gamma^{2}/9$, and $\lambda = 0.3$%
, using a generic planar initial condition. Figure \ref{fig:invariant} shows
the new invariant \eqref{eq:I_HV_DD=-4_ANO} along the trajectory (left
panel) and the relative deviation $|I(t)-I(0)|/|I(0)|$ on a logarithmic
scale (right panel). The invariant is conserved to machine precision
(relative deviation $\lesssim 10^{-12}$) throughout the integration,
confirming that it is an exact conservation law and not a numerical
artifact. For comparison, the right panel also shows the Bateman invariant
for the damped harmonic oscillator ($\lambda = 0$) with the same $a$ and $%
\gamma$, which is likewise conserved to machine precision. 

\begin{figure}[htbp]
\centering
\includegraphics[width=0.5\textwidth]{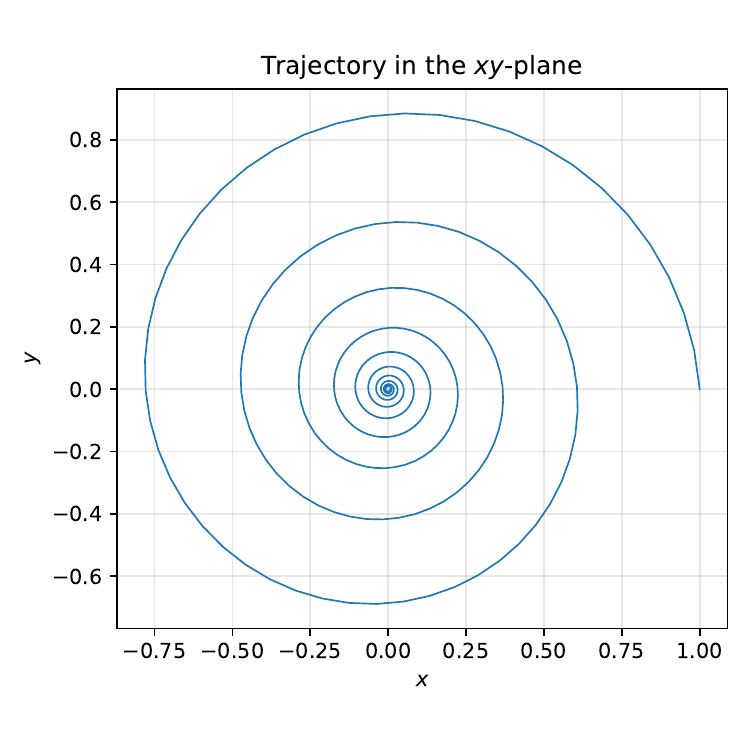}
\caption{Projection of the trajectory in the $xy$-plane for the anharmonic
DDO with $\gamma = 0.5$, $a = 2\gamma^{2}/9$, $%
\lambda = 0.3$, and the initial conditions used in Figure \ref{fig:invariant}.}
\label{fig:trajectory}
\end{figure}

\section{Conclusion}

\label{Conclusion}

In this work, we have provided, by means of Theorem \ref{Main_theorem}, an algorithm
for the classification of Noether point symmetries for time-dependent
conformal Lagrangians in an $n$-dimensional Riemannian space. There are
three quantities that one has to supply to this algorithm:

\begin{itemize}
\item[(a)] The conformal factor $A(t)$;

\item[(b)] The metric $g_{ij}$;

\item[(c)] The potential function $V(x)$.
\end{itemize}

The conformal factor fixes the time-dependent part of the (linear) friction.
The metric specifies the homothetic algebra, and the potential specifies the
damped term in the dynamical equations. Theorem \ref{Main_theorem} does not
impose any requirements or limitations on any of the three quantities, and
is therefore quite general and covariant.

We demonstrated the application of the theorem in the following
configuration:

\begin{itemize}
\item[(a)] The conformal factor $A(t) = e^{\gamma t}$;

\item[(b)] The metric is the Euclidean metric $E^{3}$;

\item[(c)] The potential is $V(x) = \frac{1}{2}ax^{i}x_{i} + \frac{\lambda}{4%
}(x^{i}x_{i})^{2}$, which defines the DO.
\end{itemize}

For this configuration Theorem \ref{Main_theorem} produced the following
results:

\begin{itemize}
\item[(a)] All the invariants of the DHO, including the classical Bateman
invariant 
\begin{equation*}
I = e^{\gamma t}\left(\dot{x}^{i}\dot{x}_{i} + \gamma x_{i}\dot{x}^{i} +
ar^{2}\right)
\end{equation*}
as a special case (Branch 2).

\item[(b)] A new exact time-dependent quadratic invariant for the anharmonic
DDO, valid when the linear stiffness satisfies $a = 2\gamma^{2}/9$: 
\begin{equation*}
I = T_{0}e^{\frac{4}{3}\gamma t}\left( x_{i}\dot{x}^{i} + \frac{3}{2\gamma}%
\dot{x}^{i}\dot{x}_{i} + \frac{\gamma}{6}r^{2} + \frac{3\lambda}{4\gamma}%
r^{4} \right).
\end{equation*}
This is the main result of the paper. Each invariant in the classification
was verified explicitly by direct differentiation along the equations of
motion, confirming the validity of Theorem \ref{Main_theorem}.

\item[(c)] The angular momentum $L_{i} = \varepsilon^{ijk}x_{j}\dot{x}_{k}$
scaled by the factor $e^{\gamma t}$, valid for all central potentials (Case
II-B).
\end{itemize}

The new invariant (b) is significant for two reasons. First, it provides a
rare example of a global conservation law for a fully nonlinear dissipative
system: for an arbitrary DDO the nonlinear coupling typically disrupts point
symmetry generators entirely, and the equation can be integrated only
through numerical or approximate means. When the linear stiffness satisfies $%
a = 2\gamma^{2}/9$, a dynamic resonance occurs in which the coordinate
dilatation vector $Y^{4}$ couples with a time-stretching transform $T(t)
\propto e^{\gamma t/3}$, producing an exact invariant. Second, this
invariant can be used for the integration of the dynamical equations of the
DDO, or as a Hamiltonian in its quantization, in the same spirit in which
the Bateman invariant is used for the DHO.

It is evident that Theorem \ref{Main_theorem} can be applied to a large
number of important nonlinear systems and produce results that would
otherwise be difficult to obtain. Future work will extend the application of
the theorem to other more general configurations, either in the conformal
factor, the metric, or the potential. One such application would be the
Cornell potential $V(x) = -\frac{a}{r} + br$ \cite%
{Eichten1975,Eichten1978,Eichten1980,CornellPotential2025}, which is the
Coulomb potential plus a linear term. A further extension, currently in
preparation, is the general case of a time-dependent separable potential $%
\omega(t)V(x)$, which opens additional avenues of application.

%\appendix

\section*{Appendix: Proof of Theorem \protect\ref{Main_theorem}}

We consider a dynamical system moving on a Riemannian manifold with metric $%
g_{ij}(x^{k})$ under the action of the time-dependent conformal Lagrangian $%
((\ln A)_{,t}\neq 0)$: 
\begin{equation}
L(t,x^{k},\dot{x}^{k})=A(t)\left[ \frac{1}{2}g_{ij}(x^{k})\dot{x}^{i}\dot{x}%
^{j}-V(x^{k})\right] .  \label{eq:GEN.1}
\end{equation}
The equation of motion of the system is 
\begin{equation}
\ddot{x}^{k}+\Gamma _{ij}^{k}\dot{x}^{i}\dot{x}^{j}+(\ln A)_{,t}\dot{x}%
^{k}+V^{,k}=0,  \label{eq:GEN.9}
\end{equation}
where $\Gamma _{ij}^{k}$ are the Christoffel symbols determined from $g_{ij}$%
.

\subsection*{The First Prolongation and Noether Condition}

A vector field $X=\xi ^{0}(t,x^{k})\partial _{t}+\xi
^{i}(t,x^{k})\partial_{x^{i}}$ generates a Noether point symmetry of $L$ if
it satisfies the Noether condition 
\begin{equation}
\mathbf{X}^{[1]}L+\frac{d\xi ^{0}}{dt}L=\frac{df}{dt},
\label{eq:NoetherCond}
\end{equation}
where $f(t,x^{k})$ is the boundary gauge function. The first prolongation $%
\mathbf{X}^{[1]}$ is given by 
\begin{equation}
\mathbf{X}^{[1]}=\xi ^{0}\partial _{t}+\xi ^{i}\partial _{x^{i}}+\left( 
\frac{d\xi ^{i}}{dt}-\dot{x}^{i}\frac{d\xi ^{0}}{dt}\right) \partial _{\dot{x%
}^{i}},
\end{equation}
where 
\begin{equation}
\frac{d\xi ^{0}}{dt}=\xi _{,t}^{0}+\xi _{,k}^{0}\dot{x}^{k},
\end{equation}
\begin{equation}
\frac{d\xi ^{i}}{dt}=\xi _{,t}^{i}+\xi _{,k}^{i}\dot{x}^{k}.
\end{equation}

We compute:%
\begin{eqnarray}
\mathbf{X}^{[1]}L &=&\frac{1}{2}\xi ^{0}\dot{A}g_{ij}\dot{x}^{i}\dot{x}^{j}
-\xi ^{0}\dot{A}V+\frac{1}{2}A\xi ^{k}g_{ij,k}\dot{x}^{i}\dot{x}^{j}
-A\xi ^{k}V_{,k}  \notag \\
&&+Ag_{mi}\xi _{,t}^{m}\dot{x}^{i}+Ag_{mi}\xi _{,j}^{m}\dot{x}^{i}\dot{x}^{j}
-Ag_{mi}\xi _{,t}^{0}\dot{x}^{m}\dot{x}^{i}
-Ag_{mi}\xi _{,j}^{0}\dot{x}^{m}\dot{x}^{i}\dot{x}^{j}.
\end{eqnarray}
\begin{equation}
L\frac{d\xi ^{0}}{dt}=\frac{1}{2}A\xi _{,t}^{0}g_{ij}\dot{x}^{i}\dot{x}^{j}
+\frac{1}{2}A\xi _{,k}^{0}g_{ij}\dot{x}^{i}\dot{x}^{j}\dot{x}^{k}
-AV\xi _{,t}^{0}-AV\xi _{,k}^{0}\dot{x}^{k}.
\end{equation}

We substitute the above into the Noether condition \eqref{eq:NoetherCond}
and collect terms with the same power of $\dot{x}^{i}$.

\subsubsection{Cubic Terms: $\dot{x}^i\dot{x}^j\dot{x}^k$}

\begin{equation}
-A(t)g_{mi}\xi _{,j}^{0}\dot{x}^{m}\dot{x}^{i}\dot{x}^{j}+\frac{1}{2}%
A(t)g_{ij}\xi _{,k}^{0}\dot{x}^{i}\dot{x}^{j}\dot{x}^{k}=0.
\end{equation}
Symmetrizing the first term (relabel $m\to i$, $i\to j$, $j\to k$) gives $%
-A(t)g_{ij}\xi _{,k}^{0}\dot{x}^{i}\dot{x}^{j}\dot{x}^{k}$, so 
\begin{equation}
-\frac{1}{2}A(t)g_{ij}\xi _{,k}^{0}\dot{x}^{i}\dot{x}^{j}\dot{x}^{k}=0,
\end{equation}
from which follows 
\begin{equation}
\xi _{,k}^{0}=0\implies \xi ^{0}=\xi ^{0}(t).  \label{eq:xi0_time_only}
\end{equation}

\subsubsection*{Quadratic Terms: $\dot{x}^i\dot{x}^j$}

Utilizing $\xi _{,k}^{0}=0$, the quadratic velocity coefficients are
collected as 
\begin{equation}
\frac{1}{2}\xi ^{0}\dot{A}g_{ij}\dot{x}^{i}\dot{x}^{j}+\frac{1}{2}%
ATY^{k}g_{ij,k}\dot{x}^{i}\dot{x}^{j}+Ag_{mi}\xi _{,j}^{m}\dot{x}^{i}\dot{x}%
^{j}-Ag_{ij}\xi _{,t}^{0}\dot{x}^{i}\dot{x}^{j}+\frac{1}{2}A\xi
_{,t}^{0}g_{ij}\dot{x}^{i}\dot{x}^{j}=0.
\end{equation}
Factoring out $A(t)$ and using the definition of the Lie derivative of the
metric tensor along the spatial generator field $\xi ^{i}=T(t)Y^{i}(x)$, we
have 
\begin{equation}
\frac{1}{2}T(t)\mathcal{L}_{Y}g_{ij}\dot{x}^{i}\dot{x}^{j}+\left[ \frac{1}{2}%
\xi ^{0}(\ln A)_{,t}-\frac{1}{2}\xi _{,t}^{0}\right] g_{ij}\dot{x}^{i}\dot{x}%
^{j}=0.
\end{equation}
This establishes that

\begin{enumerate}
\item[(a)] $Y^{i}$ must be a homothetic vector field satisfying $\mathcal{L}%
_{Y}g_{ij}=2\psi g_{ij}$ ($\psi =0$ for a Killing vector, $\psi =1$ for a
homothetic vector).

\item[(b)] The temporal functions satisfy the constraint 
\begin{equation}
2\psi T(t)+\xi ^{0}(t)(\ln A)_{,t}-\xi _{,t}^{0}(t)=0\implies \xi
_{,t}^{0}-(\ln A)_{,t}\xi ^{0}=2\psi T(t).  \label{eq:quad_constraint}
\end{equation}
\end{enumerate}

\subsubsection*{Linear Terms: $\dot{x}^{i}$}

\begin{equation}
A(t)g_{mi}\xi _{,t}^{m}\dot{x}^{i}=f_{,i}\dot{x}^{i}\implies
A(t)T_{,t}Y_{i}=f_{,i}.  \label{eq:linear_constraint}
\end{equation}
This implies that if $T_{,t}\neq 0$ the vector $Y^{i}$ is a gradient HV. Let 
$M(x)$ be the generating function so that $Y^{i}=M^{,i}$.

\subsubsection*{Zeroth-Order Terms: $(\dot{x}^i)^0$}

\begin{equation}
-\xi ^{0}\dot{A}V-ATY^{k}V_{,k}-A\xi _{,t}^{0}V=f_{,t}.  \label{eq:f_t}
\end{equation}
Equivalently, 
\begin{equation}
A(t)T(t)Y(V)+\left[ A(t)\xi ^{0}(t)\right] _{,t}V=-f_{,t}.
\label{eq:zeroth_constraint}
\end{equation}

\subsection*{The Integrability Condition for $f$}

We consider the integrability condition $f_{,it}=f_{,ti}$. Taking the
spatial derivative of equation \eqref{eq:zeroth_constraint} gives 
\begin{equation}
AT(Y(V))_{,i}+(A\xi ^{0})_{,t}V_{,i}=-f_{,ti}.
\end{equation}
Substituting the time derivative of \eqref{eq:linear_constraint}, $%
f_{,it}=(AT_{,t})_{,t}Y_{i}$, we obtain 
\begin{equation}
AT(Y(V))_{,i}+(A\xi ^{0})_{,t}V_{,i}+(AT_{,t})_{,t}Y_{i}=0.
\end{equation}
Dividing by $A(t)T(t)$ we obtain 
\begin{equation}
(Y(V))_{,i}+\frac{(A\xi ^{0})_{,t}}{AT}V_{,i}+\frac{(AT_{,t})_{,t}}{AT}%
Y_{i}=0.
\end{equation}
Since the first term depends only on the spatial coordinates, the
coefficients of the second and third terms must reduce to invariant
separation constants, which we denote as $D$ and $m$: 
\begin{equation}
\frac{(A(t)\xi ^{0}(t))_{,t}}{A(t)T(t)}=D,  \label{eq:def_D}
\end{equation}
\begin{equation}
\frac{(A(t)T_{,t}(t))_{,t}}{A(t)T(t)}=m.  \label{eq:def_m}
\end{equation}
This leads to the spatial identity governing the potential: 
\begin{equation}
(Y(V))_{,i}+DV_{,i}+mY_{i}=0.  \label{eq:constr_V}
\end{equation}
The Noether invariant is 
\begin{equation}
I=\xi ^{0}L+\left( \xi ^{i}-\xi ^{0}\dot{x}^{i}\right) \frac{\partial L}{%
\partial \dot{x}^{i}}-f(t,x).  \label{eq:Noether_invariant}
\end{equation}

\subsection*{The Constraint Equation}

Expanding the derivative in the separation definition for $m$ %
\eqref{eq:def_m} gives 
\begin{equation}
\frac{A_{,t}T_{,t}+AT_{,tt}}{AT}=m\implies T_{,tt}+(\ln
A)_{,t}T_{,t}-mT(t)=0.  \label{eq:master_T_eq}
\end{equation}
To determine the temporal component $\xi ^{0}(t)$, we expand equation %
\eqref{eq:def_D}: 
\begin{equation}
\frac{A_{,t}\xi ^{0}+A\xi _{,t}^{0}}{AT}=D\implies \xi _{,t}^{0}+(\ln
A)_{,t}\xi ^{0}=DT.  \label{xi0}
\end{equation}
We now have two independent expressions for $\xi _{,t}^{0}$: the relation
from the quadratic velocity collection \eqref{eq:quad_constraint} and this
separation identity. Equating them eliminates $\xi _{,t}^{0}$ (recall that $%
(\ln A)_{,t}\neq 0$): 
\begin{equation}
2\psi T+(\ln A)_{,t}\xi ^{0}=DT-(\ln A)_{,t}\xi ^{0} \implies 2(\ln
A)_{,t}\xi ^{0}=(D-2\psi )T,
\end{equation}
hence 
\begin{equation}
\xi ^{0}(t)=\frac{(D-2\psi )T(t)}{2(\ln A)_{,t}}.  \label{eq:master_xi0_sol}
\end{equation}
We consider cases.

\subsection*{Case I: $D-2\protect\psi \neq 0$}

Substituting \eqref{eq:master_xi0_sol} back into the quadratic constraint %
\eqref{eq:quad_constraint} yields the closed, non-linear differential
constraint restricting the conformal factor $A(t)$: 
\begin{equation}
\frac{d}{dt}\left[ \frac{(D-2\psi )T(t)}{(\ln A)_{,t}}\right] -(D-2\psi
)T(t)=4\psi T(t).  \label{eq:final_constraint}
\end{equation}
This is a necessary condition for Case I to be consistent. To proceed we
consider two cases: $T_{,t}\neq 0$ and $T_{,t}=0$.

\subsubsection*{Case I-A: $T_{,t}\neq 0$}

From \eqref{eq:linear_constraint} we have that the vector field $Y_{i}$ must
be a gradient vector, allowing us to write $Y_{i}=M_{,i}(x)$. In this case %
\eqref{eq:constr_V} gives 
\begin{equation}
Y(V)+DV+mM=C_{1}.  \label{eq:constr_V1}
\end{equation}
Also equation \eqref{eq:master_T_eq} gives 
\begin{equation}
T_{,tt}+(\ln A)_{,t}T_{,t}-mT(t)=0.  \label{eq:master_T_eq1}
\end{equation}
Equation \eqref{eq:master_xi0_sol} gives $\xi ^{0}(t)\neq 0$, and %
\eqref{eq:final_constraint} remains the same.

\paragraph{The Gauge Function and Noether First Integral.}

We integrate equation \eqref{eq:linear_constraint} spatially to determine
the gauge function up to a temporal remainder $R(t)$: 
\begin{equation}
f(t,x)=A(t)T_{,t}M(x)+R(t).
\end{equation}
Differentiating this form with respect to time gives 
\begin{equation}
f_{,t}=(AT_{,t})_{,t}M(x)+R_{,t}.
\end{equation}
Substituting this back into the zeroth-order velocity equation %
\eqref{eq:zeroth_constraint} yields 
\begin{equation}
ATY(V)+(A\xi ^{0})_{,t}V=-(AT_{,t})_{,t}M(x)-R_{,t}.
\end{equation}
Utilizing the separation constant definitions \eqref{eq:def_D} and %
\eqref{eq:def_m}, this reduces to 
\begin{equation}
AT\left[ Y(V)+DV+mM\right] =-R_{,t}.
\end{equation}
Using the identity $Y(V)+DV+mM=C_{1}$ one finds 
\begin{equation}
R_{,t}=-C_{1}A(t)T(t)\implies R(t)=-C_{1}\int A(t)T(t)\,dt.
\end{equation}
Therefore 
\begin{equation}
f(t,x)=A(t)T_{,t}M(x)-C_{1}\int A(t)T(t)\,dt.  \label{f.1}
\end{equation}
The Noether invariant is given by \eqref{eq:Noether_invariant}: 
\begin{equation}
I=A(t)\left[ T(t)Y_{i}\dot{x}^{i}-\xi ^{0}(t)\left( \frac{1}{2}\dot{x}^{i}%
\dot{x}_{i}+V(x)\right) \right] -A(t)T_{,t}M(x)+C_{1}\int A(t)T(t)\,dt.
\label{I.1}
\end{equation}

\subsubsection*{Case I-B: $T_{,t}=0$}

Setting $T=a_{1}$ constant in \eqref{eq:master_T_eq} gives $m=0$. Then %
\eqref{eq:constr_V} becomes 
\begin{equation}
Y(V)+DV=C_{2}.  \label{NL.10d1}
\end{equation}
From \eqref{eq:linear_constraint} it follows that $f=f(t)$. From %
\eqref{eq:final_constraint} it follows (with $D-2\psi \neq 0$) 
\begin{equation}
\frac{d}{dt}\left[ \frac{1}{(\ln A)_{,t}}\right] =\frac{D+2\psi}{D-2\psi}.
\label{eq:constraint_IB}
\end{equation}

The Noether function $f(t)$ is computed from the relation \eqref{eq:f_t},
from which follows 
\begin{equation*}
A(t)T(t)Y(V)+\left[ A(t)\xi ^{0}(t)\right] _{,t}V=-f_{,t},
\end{equation*}
\begin{equation*}
A(t)a_{1}(C_{2}-DV)+\left[ A(t)\xi ^{0}(t)\right] _{,t}V=-f_{,t}.
\end{equation*}
This results in the two relations 
\begin{eqnarray}
A(t)a_{1}C_{2} &=&-f_{,t},  \notag \\
\left[ A(t)\xi ^{0}(t)\right] _{,t} &=&Da_{1}A(t).  \label{eq:condn_on_xi0}
\end{eqnarray}
It can be shown that the second relation is identical with \eqref{xi0}.
Therefore only the first relation survives, which gives 
\begin{equation}
f(t)=-C_{2}a_{1}\int A(t)\,dt.  \label{eq:fin the T=0}
\end{equation}
Concerning $\xi^{0}$, from \eqref{eq:master_xi0_sol} follows 
\begin{equation}
\xi ^{0}(t)=\frac{a_{1}(D-2\psi )}{2(\ln A)_{,t}}.
\label{eq:master_xi0_sol2}
\end{equation}

The Noether integral is 
\begin{equation}
I=A(t)\left[ a_{1}Y_{i}\dot{x}^{i}-\xi ^{0}(t)\left( \frac{1}{2}\dot{x}^{i}%
\dot{x}_{i}+V(x)\right) \right] +C_{2}a_{1}\int A(t)\,dt.  \label{NL.10f}
\end{equation}
It is to be noted that the constraint \eqref{eq:final_constraint} applies to
both cases $T_{,t}\neq 0$ and $T_{,t}=0$.

\subsection*{Case II: $D-2\protect\psi =0$}

If $D=2\psi$, then from \eqref{eq:master_xi0_sol} follows 
\begin{equation}
\xi ^{0}=0,  \label{xi.D=2.0}
\end{equation}
and from the constraint \eqref{eq:quad_constraint} $\psi =0$. Therefore only
the KVs survive and $D=0$.

We now consider the spatial identity \eqref{eq:constr_V}. With $D=0$ it
reads 
\begin{equation}
(Y(V))_{,i}+mY_{i}=0.  \label{eq:constr_V_D2}
\end{equation}
We distinguish two subcases according to whether $Y_{i}$ is a gradient KV or
not.

\subsubsection*{Subcase II-A: $Y^{i}$ is a gradient KV}

If $Y_{i}=M_{,i}(x)$, then \eqref{eq:constr_V_D2} integrates to 
\begin{equation}
Y(V)+mM=C_{1},  \label{eq:constr_V_D2_grad}
\end{equation}
where $C_{1}$ is a constant. The coefficient $m$ is determined from this
relation, and $T(t)$ satisfies \eqref{THM.3}. The gauge function is computed
from \eqref{f.1} and the Noether integral from \eqref{I.1}.

\subsubsection*{Subcase II-B: $Y^{i}$ is not a gradient KV}

If $Y_{i}$ is not a gradient vector, the linear constraint 
\begin{equation*}
A(t)T_{,t}Y_{i}=f_{,i}
\end{equation*}
requires $T_{,t}=0$ (since $f_{,i}$ is exact while $Y_{i}\,dx^{i}$ is not),
so that 
\begin{equation*}
T=a_{1}=\text{constant},\qquad f_{,i}=0\ \Longrightarrow \ f=f(t).
\end{equation*}
From \eqref{eq:def_m} it then follows that 
\begin{equation}
m=0.  \label{eq:m_zero_D2}
\end{equation}
The temporal equation for $T$ is trivially satisfied, and the spatial
identity \eqref{eq:constr_V_D2} reduces to 
\begin{equation}
(Y(V))_{,i}=0\quad \Longleftrightarrow \quad Y(V)=C_{2},
\label{eq:constr_V_D2_nongrad}
\end{equation}
where $C_{2}$ is a constant.

The gauge function is obtained from \eqref{eq:f_t}, which in the present
case becomes 
\begin{equation*}
A(t)a_{1}C_{2}=-f_{,t},
\end{equation*}
from which follows 
\begin{equation}
f(t)=-C_{2}a_{1}\int A(t)\,dt.  \label{eq:fin_T0_D2}
\end{equation}

In all cases the Noether integral is given by \eqref{THM.7.1}.

This completes the proof of Theorem \ref{Main_theorem}.

\end{document}